\documentclass{article}
\usepackage{graphicx} 
\usepackage{xcolor}
\usepackage{soul}
\usepackage{amsmath}
\usepackage{amsfonts}
\usepackage{amssymb}
\usepackage{authblk}
\usepackage{hyperref}
\usepackage{longtable}
\usepackage{caption}
\usepackage{lmodern}
\usepackage[left=2cm,right=2cm,top=2cm,bottom=2cm]{geometry}
\usepackage{dcolumn}
\usepackage{bm}
\title{Variational Optimization for Constructing Inverse Potentials of Proton-Proton Scattering:\\ A Phase Function Method Study}
\author{Lalit Kumar$^{1,*}$, Arushi Sharma$^{1,*}$, Anil Khachi$^{1,2}$, Ayushi Awasthi$^{1}$ and O.S.K.S Sastri$^{1,o}$}
\affil{$^{1}$Department of Physics and Astronomical Science, Central University of Himachal Pradesh, Shahpur, Dharamshala, 176215, Himachal Pradesh, Bharat(India).}
\affil{$^{2}$Department of Physics, St. Bedes College, Navbahar, Shimla, 171002, Himachal Pradesh, Bharat(India).}

\vspace{10pt}

\begin{document}

\maketitle
\begin{abstract}

\textbf{Background}: The phase-shift analysis for proton-proton scattering has been studied by various research groups using the realistic potentials to be comprised of various internal interactions based on an exchange of pions and mesons, that involves a large no of parameters.

\textbf{Purpose}: The goal of the research is to construct inverse potentials for various $\ell$-channels of proton-proton(pp)-elastic scattering using 3 parameters Morse function in combination with atomic Hulthen by utilizing Phase function method and variational optimization technique.

\textbf{Methodology}: The implementation of variational optimization begins with assigning initial values to the Morse model parameters randomly. Utilizing the Morse + Hulthen potential as input, the phase equations for various $\ell$-channels are numerically solved using RK-5 method for obtaining the simulated Scattering Phase Shift(SPS). Mean Squared error between simulated and expected SPS has been chosen as the cost function. Variational optimization proceeds iteratively by adjusting potential parameters and re-evaluating the cost function until convergence is achieved.

\textbf{Results}:
All the obtained scattering phase shifts for various $\ell$ channels have been found to converge to a mean squared error $\leq$ 0.3. The computed cross-sections matched with the experimental ones to less than 1\% for energies up to 25 $MeV$. The scattering parameters are also found to be closely matching with the experimental data.

\textbf{Conclusion}:
The inverse potentials constructed for various $\ell$-channels using Morse+atomic Hulthen are on-par with the currently available high-precision realistic potentials.

\end{abstract}

%
%
%
%
%


\section{Introduction}
Proton-proton scattering plays a pivotal role in elucidating the dynamics and stability of nuclear fusion reactions, which offer substantial promise as a clean and sustainable energy source. The Meson Field Theory, originally proposed by Yukawa, delineates meson exchange interactions between protons and neutrons. According to Yukawa's theory, particles with mass parameters spanning the electron to nucleon masses underlie the nuclear force, offering a fundamental framework for understanding atomic nucleus binding energies.

Theoretical advancements in low-energy analyses have recently been achieved, notably through enhancements in the Coulomb potential and explicit treatment of pion-exchange effects. However, integrating the Coulomb force into nuclear reaction analysis remains a persistent challenge due to its long-range nature, impeding the application of standard techniques developed for short-range interactions. To circumvent this challenge, employing a screened Coulomb interaction has emerged as a viable approach.\\

Taylor\cite{taylor1974new} introduced an all-encompassing approach to integrating Coulomb scattering by employing

\begin{center}
 $V_c^\rho (r) = \frac{\gamma}{r}\alpha^\rho (r)$
\end{center}

where the screening function $\alpha^\rho$ for a given $\rho$ must go to zero as r tends to $\infty$ and must approach 1 as $\rho$ tends to $\infty$ with r fixed. As long as $\rho$ is very large, such a potential meets the conditions of scattering theory and produces findings that are independent of both properties of screened potential, their nature/shape, and screening radius.
In our earlier works, involving proton-proton(pp)\cite{kumar2022p,sastri2022innovative,khachi2022numerical}, p-D \cite{awasthi2023real}, $\alpha-^3He$\cite{khachi20223he}, $\alpha-\alpha$[\cite{khachi2022alpha} and $\alpha-^{12}C$\cite{kumar2022phase} scattering, we have utilized \textit{erf()} based Coulomb term as utilised by Ali and Bodmer\cite{ali1966phenomenological}, Buck\cite{buck1977local} and Odsuren\cite{odsuren2017scattering}, etc. We have observed that this \textit{erf()} based Coulomb term does not tend to zero with increasing distance, as expected from the theoretical considerations proposed by Taylor\cite{taylor1974new} while considering Coulomb interaction in Phase Function Method(PFM).

The screened Coulomb potential adeptly captures screening effects via an exponential decay in the Coulomb potential \cite{cao2014scattering}. This potential, often described by a short-range screened Coulomb potential equation, exhibits a diminishing screening function as the distance approaches infinity. Notably, the atomic Hulthén potential stands as a prominent example of exponentially screened Coulomb potentials, with wide-ranging applications across various scientific domains \cite{gonul2000hamiltonian}. 

We utilized atomic Hulthen to represent the Coulomb term and effectively generated inverse potentials for $\alpha-\alpha$\cite{awasthi2023comparative} by employing five different mathematical functions to describe nuclear interactions

The Phase Function Method (PFM), pioneered by Morse and Allis\cite{PhysRev.44.269}, facilitates the determination of Scattering Phase Shifts (SPS). Subsequent developments by  Kynch\cite{kynch1952two}, Calogero\cite{calogero1967variable}, and Bavikov\cite{babikov1967phase} have refined the phase equation, offering a method that solely relies on the interaction potential, eschewing the need for wavefunction knowledge. The efficacy of PFM has been demonstrated in analyzing scattering phenomena across various nucleon-nucleon interactions, with successful applications to realistic potentials proposed by the Nijmegen groups\cite{stoks1993partial} and Argonne v18 potentials\cite{wiringa1995accurate}. We have utilized PFM very effectively within the reference potential approach to obtain the inverse potentials for all the $\ell$-channels of neutron-proton scattering\cite{khachi2023inverse} using the Morse function as a zeroth reference. The construction of these inverse potentials was made possible by our innovative algorithm\cite{sastri2022innovative, sharma2021numerical} based on the variational Monte Carlo technique, rephrased as an optimization routine, providing an alternative to least squares minimization algorithms. Utilizing the inverse potential for the $^3S_1$ ground state we have successfully determined the deuteron structure and its form factors\cite{khachi2023deuteron}. While a simple Morse function was found to be enough for constructing inverse potentials for charge-neutral scattering, the inclusion of the Coulomb term becomes inevitable for charged particle scattering. 

This paper aims to optimize model parameters, of Morse potential for nuclear interaction combined with atomic Hulthen-based screened Coulomb potential, to obtain SPS using PFM, aligning with expected data\cite{perez2016low} to explain the experimental cross-section for proton-proton interaction.

\section{Methodology}
The Morse potential provides an effective description of the interaction between two atoms in diatomic molecules. In this study, we utilize the Morse potential\cite{morse1929diatomic} to model nuclear interactions, represented as
\begin{equation}
V_{Morse}(r) = V_0 \left( e^{-2(r - r_m)/a_m} - 2e^{-(r - r_m)/a_m} \right)
\end{equation}
where the parameters $V_0$, $r_m$, and $a_m$ denote the strength of interaction, equilibrium distance of maximum attraction, and shape of the potential, respectively. It has all the interesting features that are observed in any typical scattering experiment, such as strong repulsion at short distances, and maximum attraction at an equilibrium distance $r_m$, 
followed by a quickly decaying tail at large distances. Realistic nucleon-nucleon ($N-N$) interaction potentials, such as Argonne v18\cite{wiringa1995accurate} and Reid93\cite{modarres2004lowest} for S-states, resemble the Morse function. The phenomenological Malfleit-Tjon\cite{malfliet1969solution} potential also shares a similar shape. Furthermore, the analytical solution for the Time-Independent Schr$\ddot{o}$dinger Equation (TISE) with Morse potential interaction has been derived for bound state energies $(E < 0)$, making it a suitable choice for modeling the interaction between scattering particles\cite{morse1929diatomic}.

To account for charged hadron systems, we incorporate an electromagnetic interaction alongside the nuclear potential. We consider the electromagnetic potential, known as the Hulthen potential, as the long-range component of the effective interaction. While the pure Coulomb potential theoretically extends infinitely, it practically screens out beyond a certain distance. The atomic Hulthén potential is expressed as
\begin{equation}
V_{AH}(r) = V_0 \frac{e^{-r/a}}{1 - e^{-r/a}}
\end{equation}
where $V_0$ represents the potential strength and $a$ is the screening radius. These parameters are related by the equation $V_0 a = 2K\eta$, where $K$ is the momentum energy in the lab frame and $\eta$ is the Sommerfeld parameter defined as $\eta = \alpha/(\hbar v)$, with $v$ denoting the relative velocity of reactants at large separations and $\alpha = Z_1Z_2e^2$. For $pp$ scattering, with $Z_1 = Z_2 = 1$, $m_p = 938.272046$ $MeV$/$c^2$, and $e^2 = 1.44$ $MeV fm$, we obtain $V_0 a = 0.03472$ $fm$$^{-1}$.

\subsection{Phase Function Method}

The Time Independent Schr$\ddot{o}$dinger equation (TISE) is given as
\begin{equation}
\frac{d^2{u_{\ell}(r)}}{dr^2}+\bigg[k^2-\frac{\ell(\ell+1)}{r^2}-U(r)\bigg]u_{\ell}(r) = 0
\label{Scheq}
\end{equation}
where $U(r) = V(r)/(\hbar^2/2\mu)$, $k_{c.m} = \sqrt{E_{c.m}/(\hbar^2/2\mu)}$, and $E_{c.m} = 0.5 E_{\ell ab}$. 

The Phase Function Method is an important tool in scattering studies for both local and non-local interactions\cite{talukdar1977generalized,sett1988phase}. The TISE in Eq.~\ref{Scheq} can be transformed into a first-order nonlinear Riccati equation\cite{babikov1967phase}, which directly deals with Scattering Phase Shifts (SPS) information, as follows:
\begin{equation}
\delta_{\ell}'(k,r)=-\frac{U(r)}{k}\bigg[\cos(\delta_\ell(k,r))\hat{j}_{\ell}(kr)-\sin(\delta_\ell(k,r))\hat{\eta}_{\ell}(kr)\bigg]^2
\label{PFMeqn}
\end{equation}
where $\hat{j}_{\ell}(kr)$ represents the Riccati-Bessel function and $\hat{\eta}_{\ell}(kr)$ is the Riccati-Neumann function. By substituting the expressions for different $\ell$-values of these two functions, we obtain the respective phase equations as: 

\begin{enumerate}
\item $\ell = 0$ (S-wave):
\begin{equation}
    \delta_0'(k,r)=-\frac{U(r)}{k}\sin^2[\delta_0 + kr]
\end{equation}
\item  $\ell = 1$ (P-wave):
\begin{equation}
\delta_1'(k,r)=-\frac{U(r)}{k}\left[\frac{\sin(\delta_1+kr)-(kr) \cos(\delta_1+kr)}{kr}\right]^2
\end{equation}
\item $\ell = 2$ (D-wave):
\begin{equation}
\delta_2'(k,r) = -\frac{U(r)}{k}\left[-\sin{\left(\delta_2+ kr \right)}-\frac{3 \cos{\left(\delta_2 + kr \right)}}{kr} + \frac{3 \sin{\left(\delta_2 + kr \right)}}{{(kr)}^2}\right]^2 
\end{equation}
\item For $\ell = 3$ (F-wave):

\begin{eqnarray}\nonumber
\delta_3'(k,r)=\frac{-U(r)}{k}\Bigg[&\cos{\left(\delta_3+ kr\right)}-\frac{6}{kr}\sin{\left(\delta_3+ kr\right)}-\\ \frac{15}{{(kr)}^2}\cos{\left(\delta_3+ kr\right)}
&+\frac{15}{{(kr)}^3}\sin{\left(\delta_3+kr\right)}\Bigg]^2 
\end{eqnarray}

\item For $\ell = 4$ (G-wave):
\begin{eqnarray}\nonumber
\delta_4'(k,r)=\frac{-U(r)}{k}\Bigg[&\sin{\left(\delta_4 + kr \right)} + \frac{10 \cos{\left(\delta_4 + kr \right)}}{kr}-\frac{45 \sin{\left(\delta_4 + kr \right)}}{ {(kr)}^2}\\ 
&+\frac{105 \cos{\left(\delta_4 + kr \right)}}{ {(kr)}^3}+ \frac{105 \sin{\left(\delta_4 + kr \right)}}{{(kr)}^4}\Bigg]^2
\end{eqnarray}

\item For $\ell = 5$ (H-wave):
\begin{eqnarray}\nonumber
\delta_5'(k, r)= -\frac{U(r)}{k}&\Bigg[21\left((kr)^4-60 (kr)^2+495\right) (kr) \cos (kr)+\\ \nonumber
&\left((kr)^6-210 (kr)^4+4725 (kr)^2-10395\right)\sin (kr)\cos (\delta_5)\\ \nonumber
&-21\left((kr)^4-60 (kr)^2+495\right) (kr) \sin (kr)- \\ \nonumber
&\left((kr)^6-210 (kr)^4+4725 (kr)^2-10395\right) \cos (kr)\sin (\delta_5)\Bigg]^2\\ 
&\times \frac{1}{(kr)^{10}}
\end{eqnarray}
\end{enumerate}

These equations are solved using 5th order Runge-Kutta methods by choosing the initial condition as $\delta_{\ell}(k,0) = 0$ and integrating to a large distance.

\subsection{Optimization of Morse function Model parameters}

Typically, the variational method\cite{sastri2022innovative} is employed to determine the ground state energy for a given potential. This method initiates with a trial wave function, which is then varied randomly. This iterative process continues until convergence to the ground state is achieved. Here, instead of varying the wavefunction, we opt to vary the potential parameters and minimize the cost function to the expected data, as outlined below:

\begin{itemize}
    \item To commence the optimization procedure, the Morse parameters $V_0$, $r_m$, and $a_m$ are assigned initial values randomly from their sample space.
    \item The phase equation is integrated using the RK-5 method for different values of $k$, a function of lab energy $E$, to obtain the simulated scattering phase shift, denoted as $\delta^{sim}_k$.
    \item Define a cost function that needs to be minimized. We have chosen the mean-squared error (MSE) between the two datasets given by
    \begin{equation}
        MSE = \frac{1}{N} \sum^N_{i=1} |\delta^{exp}_{sim} - \delta^{sim}_k|^2
    \end{equation}
    where $\delta^{exp}_k$ represents experimental scattering phase shifts from the Mean-Energy Partial-Wave Analysis Data (MEPWAD) of the Granada Group.
    \item A random number $p$, generated in an interval $[-I, I]$, is added to one of the parameters, for instance, $V_{0new} = V_0 + p$.
    \item Again, the phase equation is integrated with the new set of parameters to obtain a new set of simulated scattering phase shifts, denoted as $\delta^{sim-new}_k$, using which $MSE_{new}$ is determined.
    \item If $MSE_{new} < MSE_{old}$, then $V_0 = V_{0new}$,  $MSE_{min} = MSE_{new}$, else old values are retained.
    \item The final three steps are repeated for each parameter to complete one iteration, and the size of the interval is reduced after a certain number of iterations. The process continues until $MSE_{min}$ no longer changes, indicating convergence.
\end{itemize}

\subsection{Scattering parameters}

The determination of scattering parameters, specifically the scattering length `$a_0$' and the effective range `$r_0$', holds paramount importance in understanding particle interactions within potential fields. These parameters encapsulate essential information regarding the behavior of particles undergoing scattering processes.

In line with Darewych's seminal work~\cite{darewych1967morse}, the scattering phase shift `$\delta$' can be intricately linked to the incident particle momentum `$k$' and the potential parameters `$a_0$' and `$r_0$' through the following relation:
\begin{equation}
k\cot(\delta) = -\frac{1}{a_0} + 0.5r_0 k^2
\label{sp}
\end{equation}

Here, $k$ represents the momentum of the incident particles, while $\delta$ denotes the phase shift imparted by the scattering event. By harnessing experimental data and fitting them to Eq.~\ref{sp}, researchers can accurately deduce the values of $a_0$ and $r_0$, thereby unraveling profound insights into the underlying interaction potential.

A conventional methodology employed in this pursuit involves plotting $0.5k^2$ against $k\cot(\delta)$, revealing a linear relationship. The slope of this linear correlation corresponds to $r_0$, while the intercept signifies $a_0$. This graphical analysis method offers a lucid and intuitive approach to extracting scattering parameters from experimental datasets, facilitating comprehensive comprehension of particle interactions across diverse physical systems.

\subsection{Cross section}

Once the Scattering Phase Shifts (SPS) $\delta_\ell(E)$ are obtained, for each orbital angular momentum $\ell$, one can calculate the partial Scattering Cross Section (SCS) $\sigma_\ell(E)$ using the formula\cite{amsler2015nuclear}:

\begin{equation}
\sigma_l(E;S,J)=\frac{4\pi}{k^2}\sum_{S=0}^{1}\left(\sum_{J=|\ell-S|}^{|\ell+S|}(2\ell+1)\sin^2(\delta_\ell(E;S,J))\right)
\label{a}
\end{equation}

Additionally, the total Scattering Cross Section (SCS) $\sigma_T$ is given by:

\begin{equation}
\sigma_T(E;S,J)=\frac{1}{\sum_{J=|\ell-S|}^{|\ell+S|}(2J+1)}\sum_{\ell=0}^{n}\sum_{S=0}^{1}(2J+1)\sigma_\ell(E;S,J)
\label{b}
\end{equation}

Here, 'n' represents the number of $\ell$-channels data available for the scattering state. These equations allow for the calculation of partial and total scattering cross-sections based on the obtained scattering phase shifts.

\section{Results and Discussion}

There are numerous potential models used to analyze scattering data to date. These models include Argonne v18\cite{wiringa1995accurate}, Ried soft-core potential\cite{modarres2004lowest}, NijmI, NijmII, and the new Bonn pp potential \cite{haidenbauer1989application}. Additionally, various databases are available for nucleon-nucleon scattering. Perhaps the most widely known are ENDF/B\cite{chadwick2006endf}, JENDL\cite{shibata2002japanese} nuclear data files, Nijmegen Database\cite{stoks1993partial}, and SAID (Scattering Analysis Interactive Database) maintained by George Washington University, which is a notable resource for proton-proton scattering phase shift data\cite{arndt2003said}.

In this paper, we primarily focus on the recently developed experimental MEPWAD for scattering phase shifts by P$\grave{e}$rez \textit{et al.} of the Granada group in 2016\cite{perez2016low}. The Granada group data is the latest 3$\sigma$-self-consistent database with approximately $N = 6713$ np and pp scattering data, compiled up to 2013, compared to the 1994 Nijmegen analysis \cite{stoks1993partial}, which comprises only $N = 4301$ points. We calculated the phase shifts with $\frac{\hbar^2}{m_p}= 41.47$ $MeV$ $fm^2$ and $V_0 a = 2K \eta = 0.03472 \, \text{fm}^{-1}$. Corresponding scattering phase shifts were obtained using the model parameters at various energies.

Initially, the screening radius $'a'$ in the atomic Hulth$\grave{e}$n potential has been considered as a free parameter, and optimized parameters were obtained by integrating the phase equation to a large distance in a classically forbidden region of about 40 $fm$, ensuring the potential becomes zero at infinite $r$.

The screening effect is given by exponential decay, that is, parameter $'a'$. In the asymptotic region for $r \rightarrow \infty$, 

\begin{equation}
\frac{V_0}{1-e^{-r/a}} \approx \frac{V_0a}{r}.
\end{equation}

Moreover, the screening effects of this potential increase with a decrease in the screening parameter, especially for $\ell = 3, 4, 5$. Thus, it will lead particles of these states to be weakly bound or even unbounded. In fact, particles with such weak bounds are very easy to release.

The optimized potential parameters for the different states are given in Table \ref{Table1}. These parameters, along with equations, are utilized to compute scattering phase shifts for the pp system up to partial waves $\ell = 5$.

\begin{table}[!ht]
\centering   
\caption{Model parameters for pp scattering.}
\begin{tabular}{cccccc}
\hline
\textbf{States} & \textbf{$V_0$} ($MeV$) & \textbf{$r_m$} ($fm$) & \textbf{$a_m$} ($fm$) & \textbf{a} ($fm$) & \textbf{MSE} \\ \hline 
$^1S_0$ & 91.007 & 0.872 & 0.331 & 9 & 0.1 \\ 
$^3P_0$ & 11.934 & 1.786 & 0.647 & 3 & 0.3 \\ 
$^3P_1$ & 0.010 & 4.152 & 0.696 & 3 & 0.2 \\ 
$^3P_2$ & 149.384 & 0.01 & 0.470 & 3 & 0.3\\ 
$^1D_2$ & 131.866 & 0.01 & 0.531 & 1 & 0.04  \\
$^3F_2$ & 2.728 & 2.149 & 0.551 & 0.4 & 0.003  \\
$^3F_3$ & 0.010 & 9.273 & 2.491 & 0.4 & 0.02  \\
$^3F_4$ & 11.109 & 1.419 & 0.415 & 0.4 & 0.0002  \\
$^1G_4$ & 32.341 & 0.001 & 0.724 & 0.2 & 0.002  \\
$^3H_4$ & 20.470 & 0.108 & 0.724 & 0.2 & 0.0002  \\ \hline
\end{tabular}
\label{Table1}
\end{table}

\begin{figure}[htbp]
\centering
\includegraphics[scale=0.4]{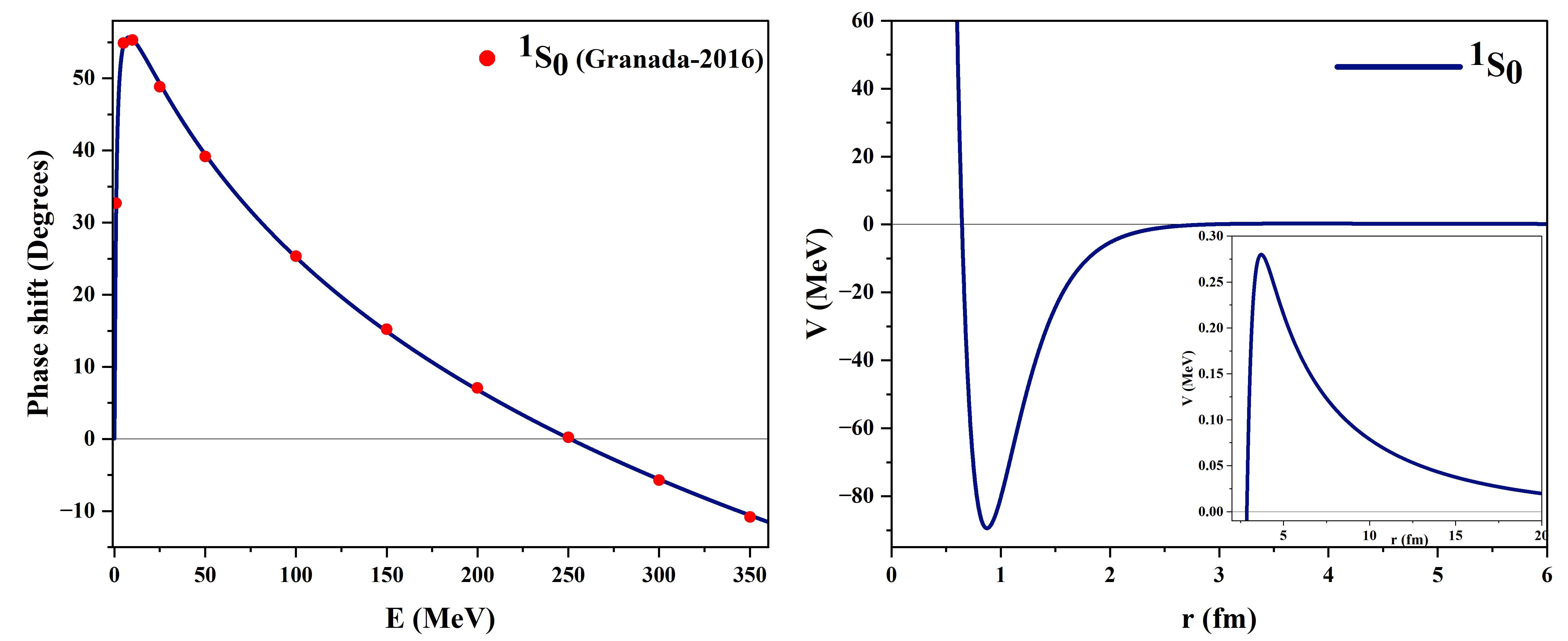}
\includegraphics[scale=0.4]{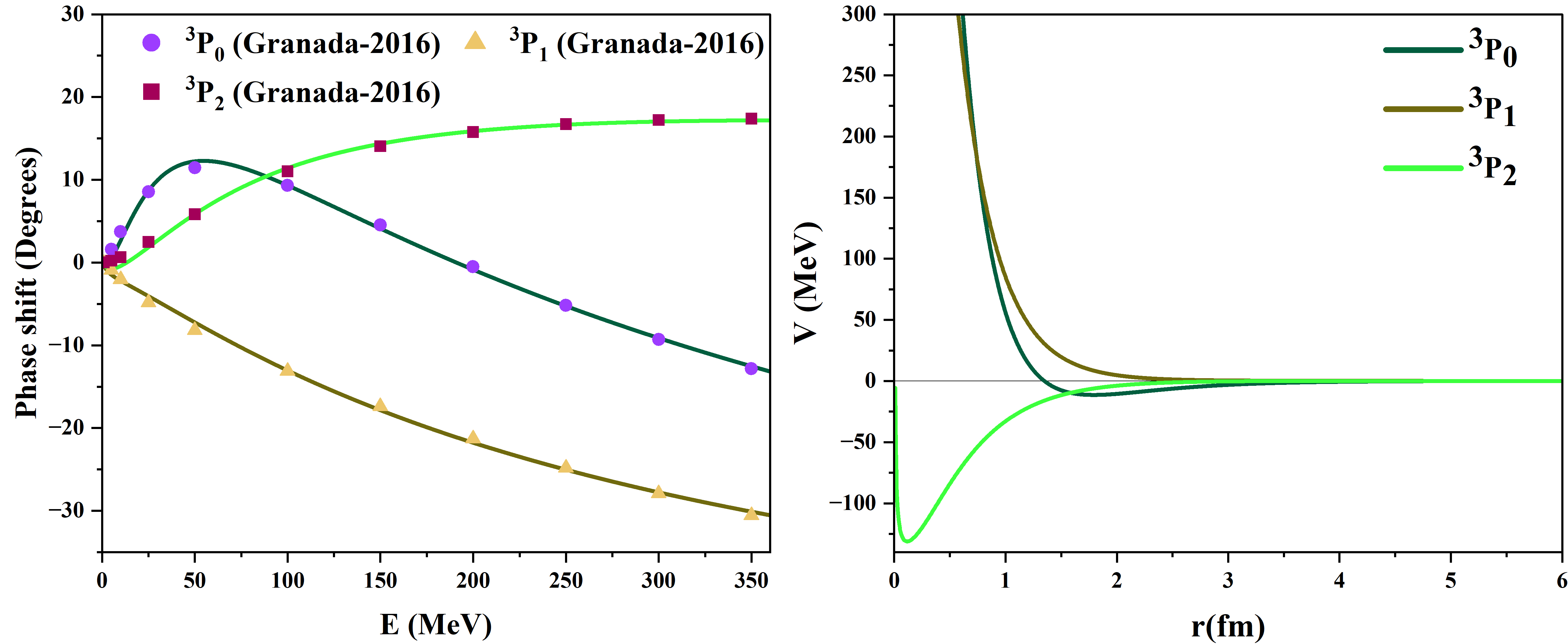}
\includegraphics[scale=0.4]{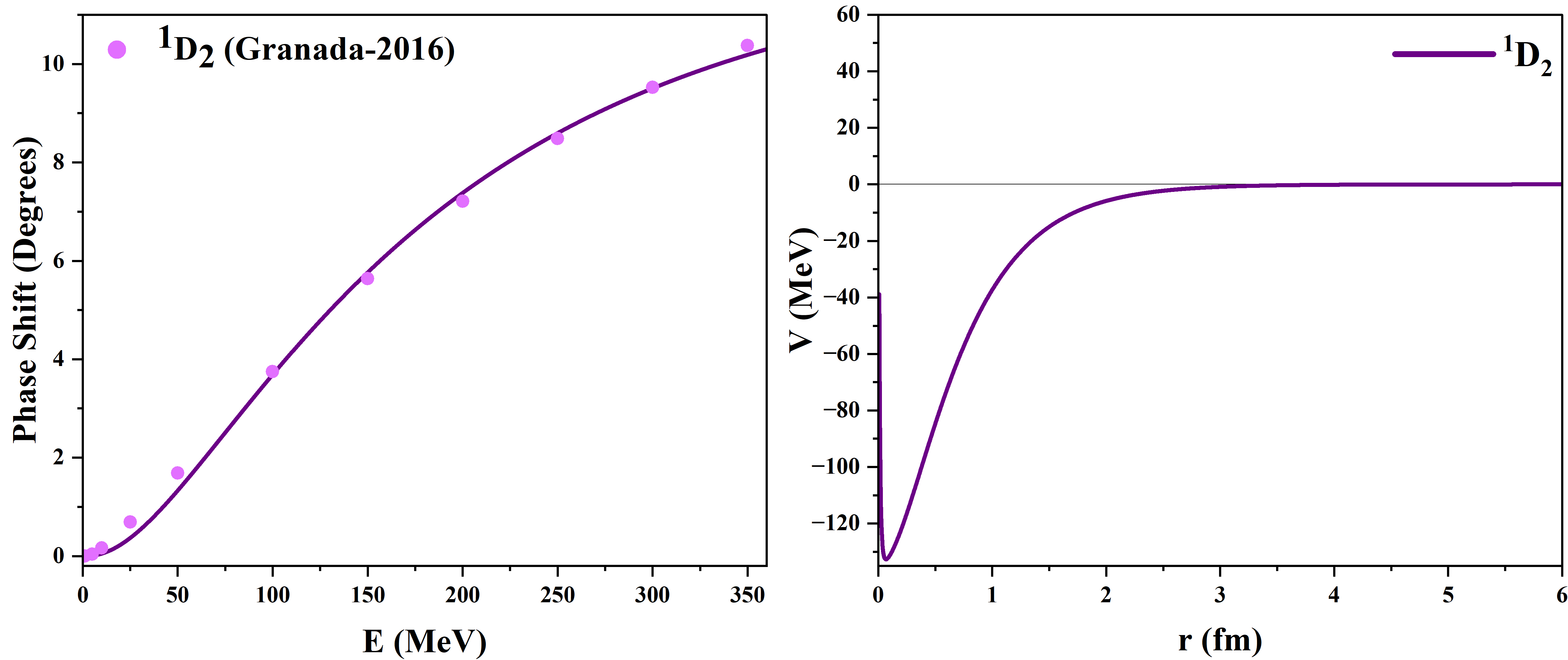}
\caption{The scattering phase shifts and inverse potentials, for the pp-interaction for S, P, and D-states are shown in the left and right columns
respectively.
}
\label{spd}
\end{figure}

From Table \ref{Table1}, it can be seen that the obtained SPS are in good match with the experimental data as the MSE values are very close to zero. In the case of $^1S_0$, the MSE comes out to be 0.1 with a Coulomb barrier of $0.280$ $MeV$, $V_0 = 91.007$ $MeV$ at an equilibrium distance of $0.872$ $fm$. fig\ref{spd} in the top right column depicts the potential for $^1S_0$, similar to one obtained with the Argonne v18 potential. In the Argonne v18 potential, authors\cite{wiringa1995accurate} have used 18 parameters to calculate potential and scattering phase shifts, whereas here, we have chosen only the Morse potential as an interaction potential, which has only 3 parameters.\\

The resulting values for the scattering length (`$a_0$') and effective range (`$r_0$') were determined(experimental~\cite{bergervoet1988phase}) to be $-7.8544(-7.8063) \text{fm}$ and $2.215(2.794) \text{fm}$, respectively, for the $^1S_0$ state. These parameters are fundamental descriptors of the scattering process, providing crucial insights into the underlying nuclear interactions and potential fields.

For $^3P_0, ^3P_1, \& ^3P_2$, the resultant MSE is 0.3, 0.2, and 0.3 with a screening parameter of $3$ $fm$. In the case of $^3P_0$ and $^3P_1$, there is a small discrepancy of 1 degree at 50 $MeV$ where the peak value does not match with the experimental ones. This may be attributed to the fact that at low energies, the Coulomb force plays a dominant role over the nuclear one. Although the phase shift results differ by a narrow margin in numerical values, they reproduce the correct trend.

\begin{figure}[htbp]
\includegraphics[scale=0.4]{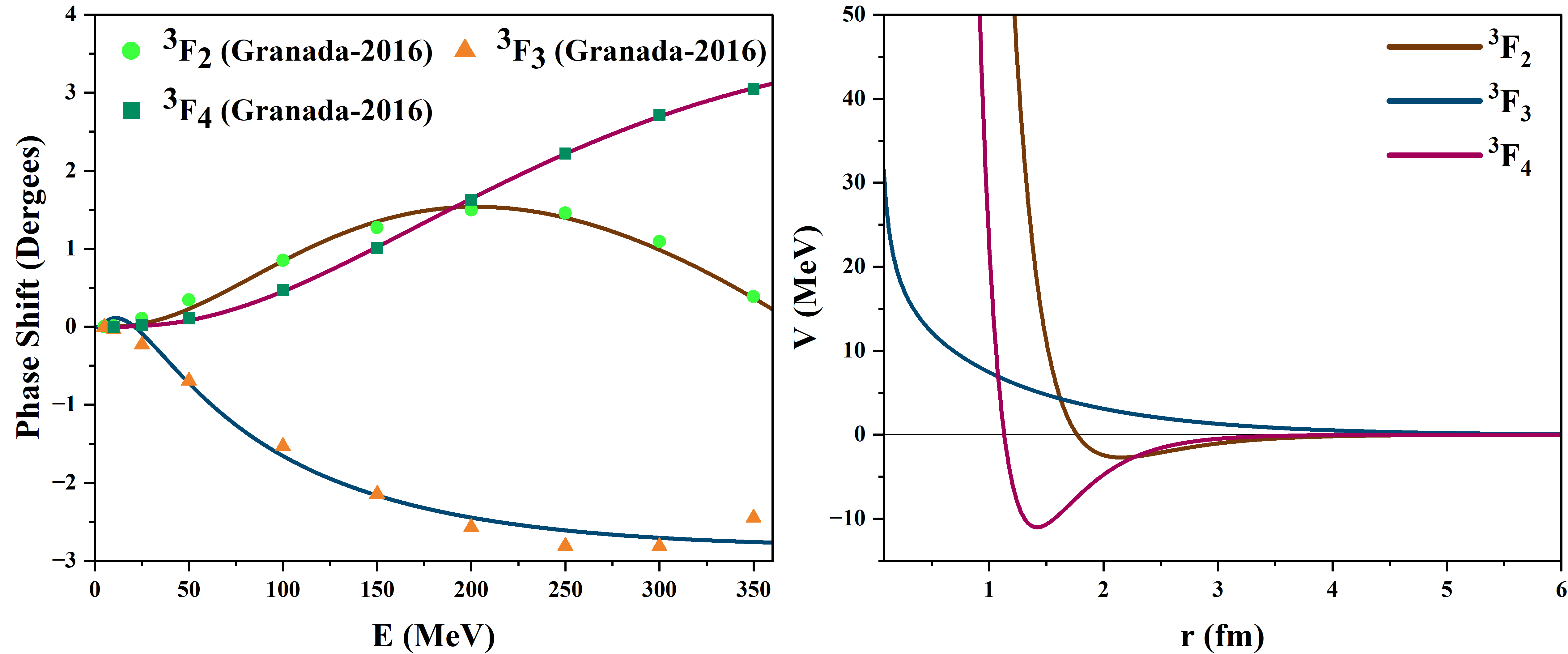}
\includegraphics[scale=0.4]{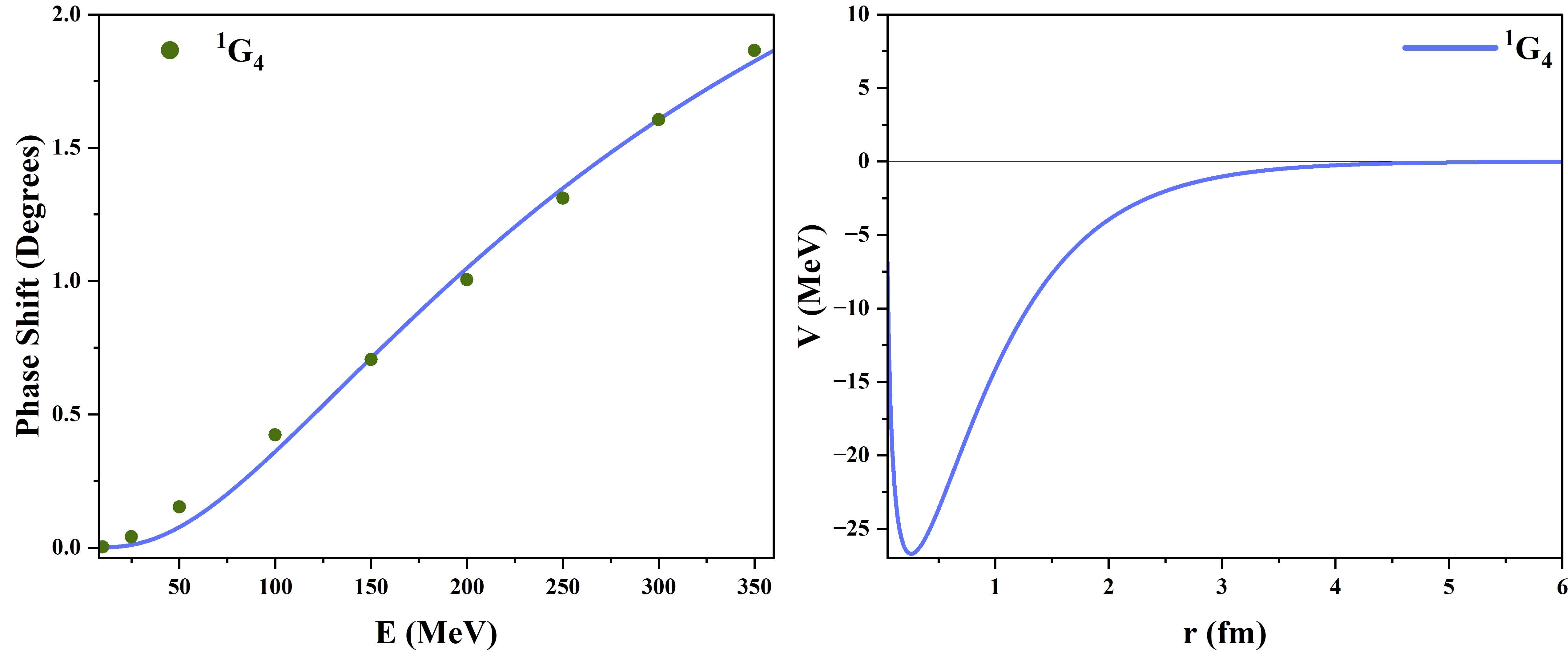}
\includegraphics[scale=0.4]{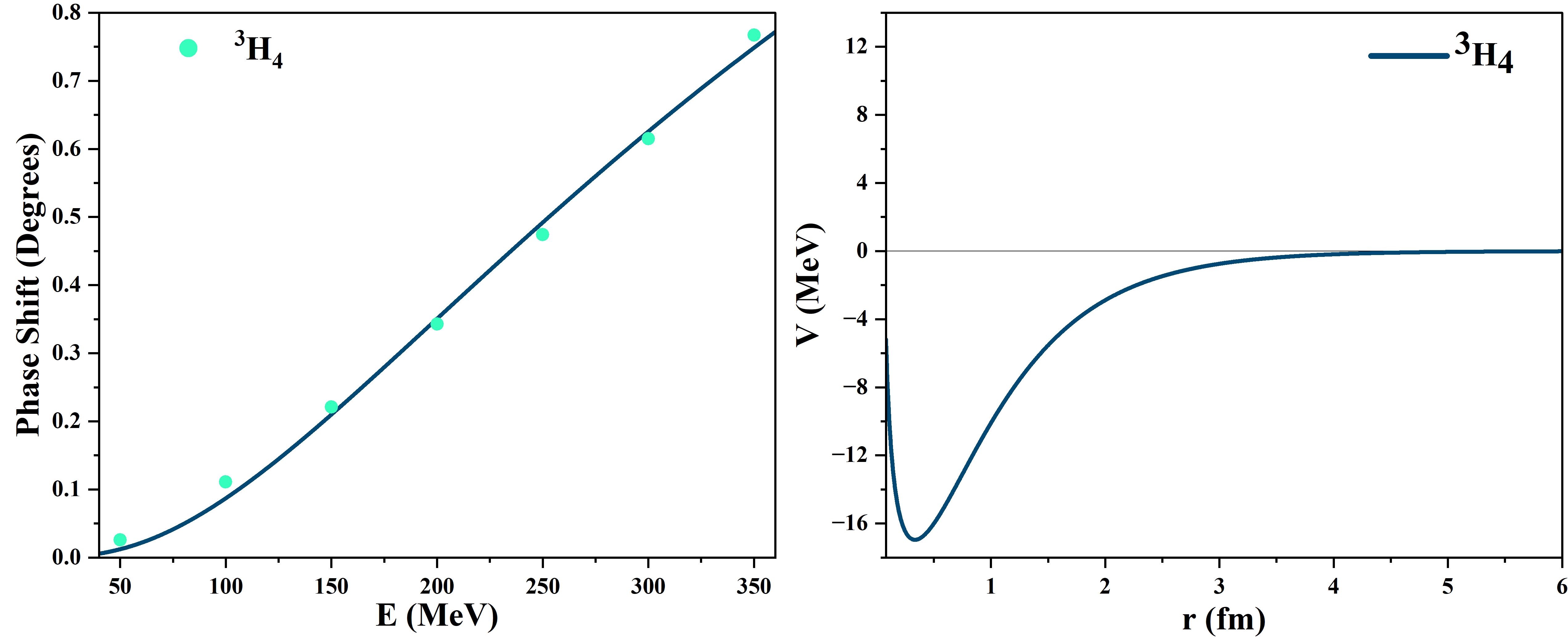}
\caption{The scattering phase shifts and inverse potentials, for the pp-interaction for F, G, and H-states are shown in the left and right columns respectively.}
\label{fgh}
\end{figure}

Similarly, fig.\ref{spd} depicts the phase shift values for the D wave and shows close agreement with the results of Ref.\cite{perez2016low} because of purely positive phase shifts, the potential is in the Gaussian form with $V_0 = 131.866$ and $r_m = 0.01$ $fm$. Similarly, for $^3F_2, ^3F_3, ^3F_4$, MSE comes out to be 0.003, 0.02, 0.002 with a screening parameter $a = 0.4$ and for $^1G_4 ~ \& ~ ^3F_4$, MSE converges to 0.002 and 0.0002 with $a = 0.2$. The difference in the numerical values is because nuclear potentials are highly state-dependent and cannot be properly generated from any known interaction, unlike atomic cases. 
The inverse potentials obtained for S, P, D, F, G, and H states are shown in the left column of fig.\ref{spd} and fig.\ref{fgh}. In addition, the scattering phase shift plots corresponding to these potentials are depicted in the right column of fig.\ref{spd} and \ref{fgh}. The following observations can be made from these potentials and SPS plots:

\begin{itemize}
\item The potentials have an attractive nature whenever the SPS are positive and a repulsive nature when the SPS are negative.
\item Whenever the SPS starts from being positive and then crosses over to negative values, as in $^1S_0$ and $^3P_0$ states, the repulsive nature of the potential curve sets in at higher values of the inter-nucleon distance.
\item It is interesting to note that in the case of the D state as well as G and H states where all the SPS are positive, the potential is in Gaussian form.
\item States with negative SPS, such as $^3P_1$ and $^3F_3$, have an exponentially decaying positive potential.
\item In the $^3F_2$ state at higher energies, the repulsive nature of the potential curve sets in at higher values of the inter-nucleon distance.
\item With increasing $\ell$, the scattering parameter goes on decreasing. That is, the screening effect of the potential goes on increasing, thus leading particles of these states to be more weakly bound.
\end{itemize}

To investigate the extent to which our model calculations yield realistic cross-section data, considering the minimal discrepancy between the results of our phase shift analysis and other calculations, we have computed the partial and total scattering cross-sections using the Eqn.\ref{a} \& \ref{b}. Notably, the $^1S_0$ state exhibits a significant contribution at low energies below 5 $MeV$, gradually decreasing with increasing energy. The obtained total scattering cross-sections closely match the experimental data\cite{arndt2009absolute}.  The total cross-section is obtained after integrating over all angles.  We have calculated both partial and total cross-sections for $\ell = 0, 1, 2, 3, 4$, and $5$ states and compared them with the reference\cite{arndt2009absolute}. The plot for the Total cross-section given in fig.\ref{partial} and the partial cross-section for all the states are given in the supplementary data. Our cross-section results for the system under consideration exhibit excellent agreement with the experimental ones. The computed values are given in supplementary material and we have observed that the obtained values match the experimental ones to less than 1$\%$ for most of the energies up to 25MeV.\\

\begin{figure}
\centering
\includegraphics[scale=0.5]{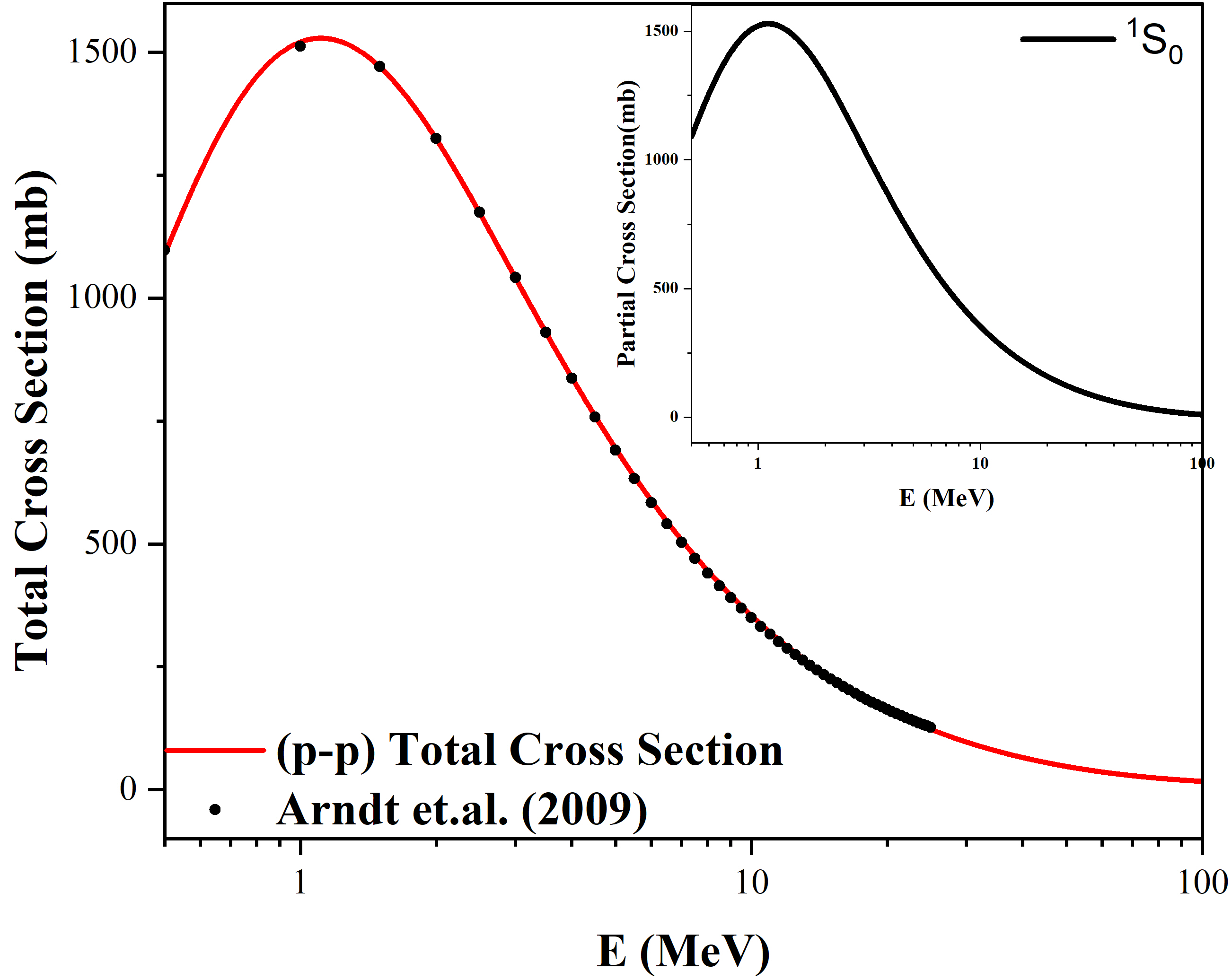}
\caption {Total Scattering Cross-section has been plotted on a log scale. Inset shows contributions due to the S-state}
\label{partial}
\end{figure}

\section{Conclusion}
\begin{itemize}
    \item In this work, we have utilized a 3-parameter Morse function along with atomic Hulthen as the zeroth reference to construct these inverse potentials for various $\ell$-channels in the energy range 1-350 $MeV$.
    \item The screening parameter in atomic Hulthen decreases with increasing $\ell$-values. This leads to the weakening or even unbinding of particles in these states, highlighting the importance of screening effects in nuclear interactions.
    \item The obtained Scattering Phase shifts with the inverse potentials for each of the $\ell$-channels match the expected ones to MSE of $\leq 0.3$.
    \item The total scattering cross-section obtained using these inverse potentials matches with experimental values to the order of 1-2\% for energies up to 25 $MeV$.
    \item One of the major advantages of these inverse potentials is their ease of portability for performing ab-initio nuclear structure calculations. The overall performance of our constructed inverse potentials for all $\ell$-channels are on-par with the high precision realistic potentials available to date.
    \item To further enhance the representation of Coulomb potential, a combination of piece-wise smooth regular Morse and inverted Morse is being attempted.
\end{itemize}
\section*{Acknowledgment}
A. Awasthi acknowledges financial support provided by Department of Science and Technology
(DST), Government of India vide Grant No. DST/INSPIRE Fellowship/2020/IF200538.   \\
 
\textbf{Author Declaration} The authors declare that they have no conflict of interest.
\section*{Data Availability}
The datasets used and/or analyzed during the current study are available from the Granada Database. \\
https://www.ugr.es/~amaro/nndatabase/

\section{Appendix}

The phase shift values derived from the inverse potentials compared to the experimental data are provided in Table \ref{tab:1} and \ref{tab:2}.

\begin{table}[htbp]
\centering
\caption{Calculated Scattering Phase shifts along with their expected data for S, P and D state}
\resizebox{1.1\textwidth}{!}{
\begin{tabular}{|c|cc|cc|cc|cc|cc|}
\hline
       & \multicolumn{2}{c|}{$^1S_0$}                               & \multicolumn{2}{c|}{$^3P_0$}                             & \multicolumn{2}{c|}{$^3P_1$}                             & \multicolumn{2}{c|}{$^3P_2$}                                         & \multicolumn{2}{c|}{$^1D_2$}                                       \\ \hline
Energy & \multicolumn{1}{c|}{DeltaExp(degrees)} & {DeltaSim(degrees)} & \multicolumn{1}{c|}{DeltaExp(degrees)} &{ DeltaSim(degrees)} & \multicolumn{1}{c|}{DeltaExp(degrees)} &{DeltaSim(degrees)} & \multicolumn{1}{c|}{DeltaExp(degrees)} & DeltaSim(degrees) & \multicolumn{1}{c|}{DeltaExp(degrees)} & DeltaSim(degrees) \\ \hline
1      & \multicolumn{1}{c|}{32.677}         & 32.688         & \multicolumn{1}{c|}{0.133}          & -0.178         & \multicolumn{1}{c|}{-0.079}         & -0.34          & \multicolumn{1}{c|}{0.014}                & -0.284               & \multicolumn{1}{c|}{0.001}              & 0                  \\ \hline
5      & \multicolumn{1}{c|}{54.895}         & 54.625         & \multicolumn{1}{c|}{1.575}          & 0.516          & \multicolumn{1}{c|}{-0.888}         & -1.369         & \multicolumn{1}{c|}{0.214}                & -0.76                & \multicolumn{1}{c|}{0.042}              & 0.008              \\ \hline
10     & \multicolumn{1}{c|}{55.32}          & 55.418         & \multicolumn{1}{c|}{3.717}          & 2.664          & \multicolumn{1}{c|}{-2.028}         & -2.074         & \multicolumn{1}{c|}{0.648}                & -0.415               & \multicolumn{1}{c|}{0.164}              & 0.042              \\ \hline
25     & \multicolumn{1}{c|}{48.848}         & 49.296         & \multicolumn{1}{c|}{8.552}          & 8.683          & \multicolumn{1}{c|}{-4.84}          & -3.917         & \multicolumn{1}{c|}{2.479}                & 1.84                 & \multicolumn{1}{c|}{0.689}              & 0.34               \\ \hline
50     & \multicolumn{1}{c|}{39.182}         & 39.461         & \multicolumn{1}{c|}{11.436}         & 12.222         & \multicolumn{1}{c|}{-8.161}         & -7.123         & \multicolumn{1}{c|}{5.837}                & 5.846                & \multicolumn{1}{c|}{1.685}              & 1.299              \\ \hline
100    & \multicolumn{1}{c|}{25.357}         & 25.203         & \multicolumn{1}{c|}{9.324}          & 9.313          & \multicolumn{1}{c|}{-13.109}        & -13.036        & \multicolumn{1}{c|}{11.027}               & 11.407               & \multicolumn{1}{c|}{3.75}               & 3.686              \\ \hline
150    & \multicolumn{1}{c|}{15.229}         & 14.914         & \multicolumn{1}{c|}{4.532}          & 4.106          & \multicolumn{1}{c|}{-17.379}        & -17.882        & \multicolumn{1}{c|}{14.059}               & 14.342               & \multicolumn{1}{c|}{5.639}              & 5.784              \\ \hline
200    & \multicolumn{1}{c|}{7.076}          & 6.819          & \multicolumn{1}{c|}{-0.494}         & -0.857         & \multicolumn{1}{c|}{-21.273}        & -21.822        & \multicolumn{1}{c|}{15.768}               & 15.868               & \multicolumn{1}{c|}{7.212}              & 7.399              \\ \hline
250    & \multicolumn{1}{c|}{0.212}          & 0.126          & \multicolumn{1}{c|}{-5.16}          & -5.259         & \multicolumn{1}{c|}{-24.784}        & -25.065        & \multicolumn{1}{c|}{16.721}               & 16.653               & \multicolumn{1}{c|}{8.487}              & 8.605              \\ \hline
300    & \multicolumn{1}{c|}{-5.694}         & -5.589         & \multicolumn{1}{c|}{-9.287}         & -9.125         & \multicolumn{1}{c|}{-27.88}         & -27.77         & \multicolumn{1}{c|}{17.208}               & 17.031               & \multicolumn{1}{c|}{9.525}              & 9.499              \\ \hline
350    & \multicolumn{1}{c|}{-10.828}        & -10.581        & \multicolumn{1}{c|}{-12.813}        & -12.529        & \multicolumn{1}{c|}{-30.527}        & -30.053        & \multicolumn{1}{c|}{17.376}               & 17.174               & \multicolumn{1}{c|}{10.375}             & 10.163             \\ \hline
\end{tabular}}

\label{tab:1}
\end{table}

\begin{table}[htbp]
\centering
\caption{Calculated Scattering Phase-shift along with their expected values for F, G and H state}
\resizebox{1.1\textwidth}{!}{
\begin{tabular}{|c|cc|cc|cc|cc|cc|}
\hline
            & \multicolumn{2}{c|}{$^3F_2$}                                   & \multicolumn{2}{c|}{$^3F_3$}                                   & \multicolumn{2}{c|}{$^3F_4$}                                   & \multicolumn{2}{c|}{$^1G_4$}                                     & \multicolumn{2}{c|}{$^3H_4$}                                     \\ \hline
\#Elab($MeV$) & \multicolumn{1}{c|}{DeltaExp(degrees)} & DeltaSim(degrees) & \multicolumn{1}{c|}{DeltaExp(degrees)} & DeltaSim(degrees) & \multicolumn{1}{c|}{DeltaExp(degrees)} & DeltaSim(degrees) & \multicolumn{1}{c|}{DeltaExp(degrees)} & DeltaSim(degrees) & \multicolumn{1}{c|}{DeltaExp(degrees)} & DeltaSim(degrees) \\ \hline
5           & \multicolumn{1}{c|}{0.002}             & 0                 & \multicolumn{1}{c|}{-0.004}            & 0.052             & \multicolumn{1}{c|}{}                  &                   & \multicolumn{1}{c|}{}                  &                   & \multicolumn{1}{c|}{}                  &                   \\ \hline
10          & \multicolumn{1}{c|}{0.013}             & 0.003             & \multicolumn{1}{c|}{-0.031}            & 0.111             & \multicolumn{1}{c|}{0.001}             & 0.001             & \multicolumn{1}{c|}{0.003}             & 0                 & \multicolumn{1}{c|}{}                  &                   \\ \hline
25          & \multicolumn{1}{c|}{0.106}             & 0.039             & \multicolumn{1}{c|}{-0.231}            & -0.085            & \multicolumn{1}{c|}{0.02}              & 0.011             & \multicolumn{1}{c|}{0.04}              & 0.01              & \multicolumn{1}{c|}{0.004}             & 0.001             \\ \hline
50          & \multicolumn{1}{c|}{0.345}             & 0.225             & \multicolumn{1}{c|}{-0.692}            & -0.723            & \multicolumn{1}{c|}{0.108}             & 0.083             & \multicolumn{1}{c|}{0.153}             & 0.077             & \multicolumn{1}{c|}{0.026}             & 0.012             \\ \hline
100         & \multicolumn{1}{c|}{0.853}             & 0.834             & \multicolumn{1}{c|}{-1.53}             & -1.657            & \multicolumn{1}{c|}{0.471}             & 0.454             & \multicolumn{1}{c|}{0.423}             & 0.36              & \multicolumn{1}{c|}{0.111}             & 0.087             \\ \hline
150         & \multicolumn{1}{c|}{1.271}             & 1.339             & \multicolumn{1}{c|}{-2.142}            & -2.163            & \multicolumn{1}{c|}{1.011}             & 1.023             & \multicolumn{1}{c|}{0.706}             & 0.711             & \multicolumn{1}{c|}{0.221}             & 0.21              \\ \hline
200         & \multicolumn{1}{c|}{1.499}             & 1.536             & \multicolumn{1}{c|}{-2.568}            & -2.446            & \multicolumn{1}{c|}{1.628}             & 1.642             & \multicolumn{1}{c|}{1.005}             & 1.049             & \multicolumn{1}{c|}{0.343}             & 0.351             \\ \hline
250         & \multicolumn{1}{c|}{1.457}             & 1.415             & \multicolumn{1}{c|}{-2.809}            & -2.61             & \multicolumn{1}{c|}{2.22}              & 2.216             & \multicolumn{1}{c|}{1.311}             & 1.348             & \multicolumn{1}{c|}{0.474}             & 0.492             \\ \hline
300         & \multicolumn{1}{c|}{1.093}             & 1.029             & \multicolumn{1}{c|}{-2.814}            & -2.705            & \multicolumn{1}{c|}{2.71}              & 2.694             & \multicolumn{1}{c|}{1.605}             & 1.606             & \multicolumn{1}{c|}{0.615}             & 0.626             \\ \hline
350         & \multicolumn{1}{c|}{0.389}             & 0.439             & \multicolumn{1}{c|}{-2.447}            & -2.759            & \multicolumn{1}{c|}{3.048}             & 3.057             & \multicolumn{1}{c|}{1.865}             & 1.825             & \multicolumn{1}{c|}{0.767}             & 0.749             \\ \hline
\end{tabular}}

\label{tab:2}
\end{table}

\begin{longtable}{|c|c|c|c|c|c|c|c|c|}
\caption{Comparison of Total Cross Sections for S, P, D, F, G, and H States with Experimental data} \label{tab:3}\\
\hline
E($MeV$) & $\sigma_{exp}$(barn)   & $\sigma_s$      & $\sigma_p$     & $\sigma_d$        & $\sigma_f$        & $\sigma_g$       & $\sigma_h$  & $\sigma_{Total}$ (barn) \\ 
\hline
\endhead 
\hline
0.5    & 1.098  & 1.0887 & 2E-04 & 8.16E-12 & 0        & 0        & 0 & 1.0889      \\ \hline
1      & 1.513  & 1.5210 & 5E-04 & 1.34E-10 & 0        & 0        & 0 & 1.5215      \\ \hline
1.5    & 1.471  & 1.4700 & 7E-04 & 6.89E-10 & 0        & 0        & 0 & 1.4707      \\ \hline
2      & 1.325  & 1.3233 & 9E-04 & 2.21E-09 & 0        & 0        & 0 & 1.3241      \\ \hline
2.5    & 1.175  & 1.1720 & 1E-03 & 5.47E-09 & 0        & 0        & 0 & 1.1729      \\ \hline
3      & 1.042  & 1.0400 & 0.001 & 1.15E-08 & 0        & 0        & 0 & 1.0410      \\ \hline
3.5    & 0.9306 & 0.9296 & 0.001 & 2.14E-08 & 0        & 0        & 0 & 0.9307      \\ \hline
4      & 0.8367 & 0.8375 & 0.001 & 3.67E-08 & 0        & 0        & 0 & 0.8385      \\ \hline
4.5    & 0.7577 & 0.7597 & 0.001 & 5.91E-08 & 0        & 0        & 0 & 0.7607      \\ \hline
5      & 0.6906 & 0.6934 & 0.001 & 9.03E-08 & 2.23E-06 & 0        & 0 & 0.6945      \\ \hline
5.5    & 0.6333 & 0.6366 & 0.001 & 1.32E-07 & 2.79E-06 & 0        & 0 & 0.6376      \\ \hline
6      & 0.5838 & 0.5876 & 0.001 & 1.87E-07 & 3.33E-06 & 0        & 0 & 0.5886      \\ \hline
6.5    & 0.5408 & 0.5449 & 0.001 & 2.58E-07 & 3.82E-06 & 0        & 0 & 0.5460      \\ \hline
7      & 0.5031 & 0.5075 & 0.001 & 3.46E-07 & 4.24E-06 & 0        & 0 & 0.5086      \\ \hline
7.5    & 0.4699 & 0.4744 & 0.001 & 4.54E-07 & 4.58E-06 & 0        & 0 & 0.4755      \\ \hline
8      & 0.4404 & 0.4449 & 0.001 & 5.84E-07 & 4.84E-06 & 0        & 0 & 0.4460      \\ \hline
8.5    & 0.414  & 0.4185 & 0.001 & 7.40E-07 & 5.00E-06 & 0        & 0 & 0.4196      \\ \hline
9      & 0.3903 & 0.3947 & 0.001 & 9.25E-07 & 5.08E-06 & 0        & 0 & 0.3958      \\ \hline
9.5    & 0.369  & 0.3731 & 0.001 & 1.14E-06 & 5.08E-06 & 0        & 0 & 0.3742      \\ \hline
10     & 0.3496 & 0.3535 & 0.001 & 1.39E-06 & 5.01E-06 & 0        & 0 & 0.3547      \\ \hline
10.5   & 0.332  & 0.3356 & 0.001 & 1.67E-06 & 4.88E-06 & 7.75E-10 & 0 & 0.3369      \\ \hline
11     & 0.316  & 0.3193 & 0.001 & 2.00E-06 & 4.70E-06 & 1.27E-09 & 0 & 0.3206      \\ \hline
11.5   & 0.3012 & 0.3043 & 0.001 & 2.36E-06 & 4.48E-06 & 1.79E-09 & 0 & 0.3056      \\ \hline
12     & 0.2877 & 0.2905 & 0.001 & 2.77E-06 & 4.22E-06 & 2.32E-09 & 0 & 0.2918      \\ \hline
12.5   & 0.2752 & 0.2776 & 0.001 & 3.23E-06 & 3.94E-06 & 3.02E-09 & 0 & 0.2791      \\ \hline
13     & 0.2637 & 0.2657 & 0.001 & 3.74E-06 & 3.64E-06 & 3.71E-09 & 0 & 0.2672      \\ \hline
13.5   & 0.253  & 0.2547 & 0.002 & 4.30E-06 & 3.32E-06 & 4.56E-09 & 0 & 0.2562      \\ \hline
14     & 0.2431 & 0.2444 & 0.002 & 4.91E-06 & 3.00E-06 & 5.51E-09 & 0 & 0.2460      \\ \hline
14.5   & 0.2338 & 0.2347 & 0.002 & 5.59E-06 & 2.68E-06 & 6.60E-09 & 0 & 0.2364      \\ \hline
15     & 0.2252 & 0.2257 & 0.002 & 6.32E-06 & 2.36E-06 & 7.85E-09 & 0 & 0.2274      \\ \hline
15.5   & 0.2171 & 0.2173 & 0.002 & 7.11E-06 & 2.04E-06 & 9.29E-09 & 0 & 0.2190      \\ \hline
16     & 0.2096 & 0.2093 & 0.002 & 7.97E-06 & 1.74E-06 & 1.09E-08 & 0 & 0.2112      \\ \hline
16.5   & 0.2025 & 0.2018 & 0.002 & 8.90E-06 & 1.45E-06 & 1.28E-08 & 0 & 0.2038      \\ \hline
17     & 0.1958 & 0.1948 & 0.002 & 9.90E-06 & 1.18E-06 & 1.49E-08 & 0 & 0.1968      \\ \hline
17.5   & 0.1895 & 0.1882 & 0.002 & 1.10E-05 & 9.36E-07 & 1.72E-08 & 0 & 0.1902      \\ \hline
18     & 0.1836 & 0.1819 & 0.002 & 1.21E-05 & 7.15E-07 & 1.99E-08 & 0 & 0.1840      \\ \hline
18.5   & 0.178  & 0.1759 & 0.002 & 1.33E-05 & 5.23E-07 & 2.28E-08 & 0 & 0.1781      \\ \hline
19     & 0.1728 & 0.1702 & 0.002 & 1.46E-05 & 3.61E-07 & 2.61E-08 & 0 & 0.1725      \\ \hline
19.5   & 0.1678 & 0.1649 & 0.002 & 1.60E-05 & 2.32E-07 & 2.98E-08 & 0 & 0.1672      \\ \hline
20     & 0.163  & 0.1597 & 0.002 & 1.74E-05 & 1.38E-07 & 3.38E-08 & 0 & 0.1621      \\ \hline
20.5   & 0.1587 & 0.1549 & 0.002 & 1.90E-05 & 8.12E-08 & 3.82E-08 & 0 & 0.1573      \\ \hline
21     & 0.1544 & 0.1502 & 0.002 & 2.06E-05 & 6.18E-08 & 4.30E-08 & 0 & 0.1528      \\ \hline
21.5   & 0.1504 & 0.1458 & 0.003 & 2.23E-05 & 8.12E-08 & 4.83E-08 & 0 & 0.1484      \\ \hline
22     & 0.1465 & 0.1416 & 0.003 & 2.41E-05 & 1.40E-07 & 5.41E-08 & 0 & 0.1443      \\ \hline
22.5   & 0.1428 & 0.1376 & 0.003 & 2.59E-05 & 2.39E-07 & 6.04E-08 & 0 & 0.1403      \\ \hline
23     & 0.1393 & 0.1337 & 0.003 & 2.79E-05 & 3.77E-07 & 6.73E-08 & 0 & 0.1365      \\ \hline
23.5   & 0.1359 & 0.1301 & 0.003 & 2.99E-05 & 5.56E-07 & 7.47E-08 & 0 & 0.1329      \\ \hline
24     & 0.1327 & 0.1265 & 0.003 & 3.21E-05 & 7.74E-07 & 8.28E-08 & 0 & 0.1294      \\ \hline
24.5   & 0.1296 & 0.1231 & 0.003 & 3.43E-05 & 1.03E-06 & 9.14E-08 & 0 & 0.1261      \\ \hline
25     & 0.1267 & 0.1199 & 0.003 & 3.66E-05 & 1.33E-06 & 1.01E-07 & 0 & 0.1229      \\ \hline


\end{longtable}

\begin{table}[htbp]
    \centering
     \caption{The individual contributions to the calculated total elastic scattering cross-section (SCS) from various channels.}
    \resizebox{1.1\textwidth}{!}{
    \begin{tabular}{|l|l|l|l|l|l|l|l|l|}
    \hline
        E($MeV$) & $\sigma_s$ & $\sigma_p$ & $\sigma_d$ & $\sigma_f$ & $\sigma_g$ & $\sigma_h$ & $\sigma_{sim}$& $\sigma_{exp}$ \\ \hline
        1 & 1.521(99.97\%) & 0.00046(0.03\%) & 0 & 0 & 0 & 0 & 1.5215 & 1.513 \\ \hline
        10 & 0.3535(99.96\%) & 0.00119(0.34\%) & 0 & 0 & 0 & 0 & 0.3547 & 0.3496 \\ \hline
        50 & 0.0421(88.77\%) & 0.00502(10.57\%) & 0.0003(0.56\%) & 4.22$\times 10^{-5}$(0.1\%) & 0 & 0 & 0.0475 & 0.05895 \\ \hline
        100 & 0.0095(54.97\%) & 0.00651(37.85\%) & 0.0011(6.27\%) & 0.0001(0.79\%) & 0.00002(0.12\%) & 0 & 0.0172 & 0.0317 \\ \hline
        150 & 0.0023(20.59\%) & 0.00688(61.47\%) & 0.0018(15.78\%) & 0.0002(1.66\%) & 0.00005(0.45\%) & 0.0000052(0.05\%) & 0.0112 & 0.02574 \\ \hline
        200 & 0.0004(3.8\%) & 0.00684(70.68\%) & 0.0022(22.36\%) & 0.0002(2.21\%) & 0.0001(0.83\%) & 0.0000108(0.11\%) & 0.0097 & 0.02422 \\ \hline
        250 & 0 & 0.00665(71.22\%) & 0.0023(25.02\%) & 0.0002(2.45\%) & 0.0001(1.13\%) & 0.000017(0.18\%) & 0.0093 & 0.02389 \\ \hline
        300 & 0.0002(1.77\%) & 0.0064(68.75\%) & 0.0024(25.41\%) & 0.0002(2.5\%) & 0.0001(1.33\%) & 0.0000229(0.24\%) & 0.0093 & 0.02407 \\ \hline
        350 & 0.0005(5.37\%) & 0.00615(65.64\%) & 0.0023(24.77\%) & 0.0002(2.46\%) & 0.0001(1.46\%) & 0.000028(0.3\%) & 0.0094 & 0.02493 \\ \hline
    \end{tabular}}
    
    \label{tab:4}
\end{table}

Tables \ref{tab:3} and \ref{tab:4} illustrate the partial cross-section contributions from various angular momentum states ($\ell$-states) to the overall cross-section. To compare with the experimental cross-section provided by Arndt, total cross-sections up to 25 $MeV$ are given in Table \ref{tab:3}. Table \ref{tab:4} extends the cross-section values up to 350 $MeV$. The percentage contributions made by the S-state and the other $\ell$-channels from P to H to the estimated total Cross Section are shown in brackets. Analysis of Table \ref{tab:4} reveals that at lower energies, the primary contribution comes from the s-state, while higher $\ell$-states become increasingly significant with increasing energy. The influence of P and D channels becomes notable at higher energy levels, spanning from 100 $MeV$ to 350 $MeV$. The contributions from the F and G states are notably lower within the given range, yet they become increasingly significant for precisely characterizing the observed experimental total cross-section. However, the SPS is only accessible for one H-state, and its impact on determining the total scattering cross-section is nearly insignificant.

\begin{figure}[htbp]
    \centering
    \includegraphics[scale=0.35]{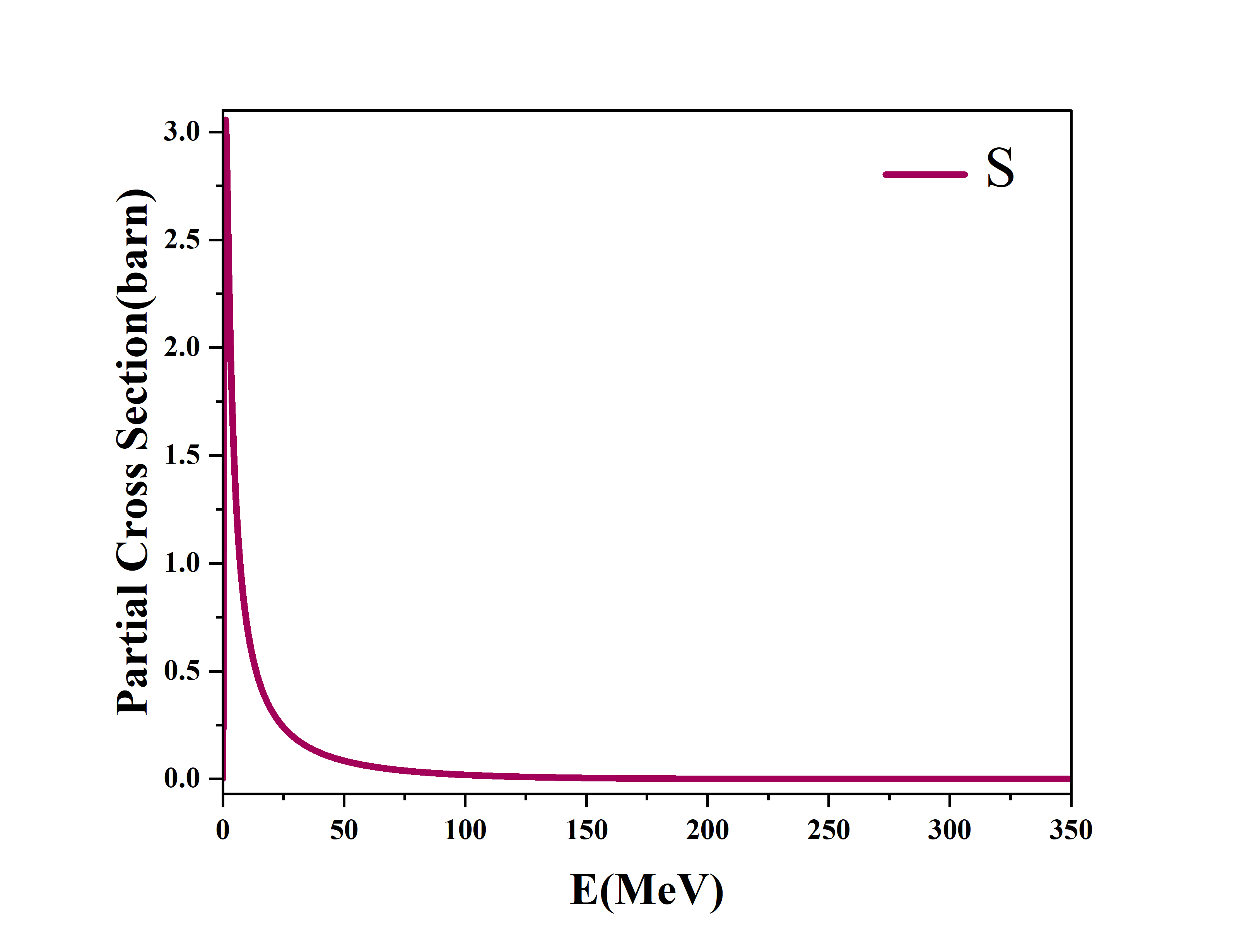}
    \caption{Partial Cross section for S state as a function of energy}
    \label{fig:s}
\end{figure}

\begin{figure}[htbp]
    \centering
    \includegraphics[scale=0.35]{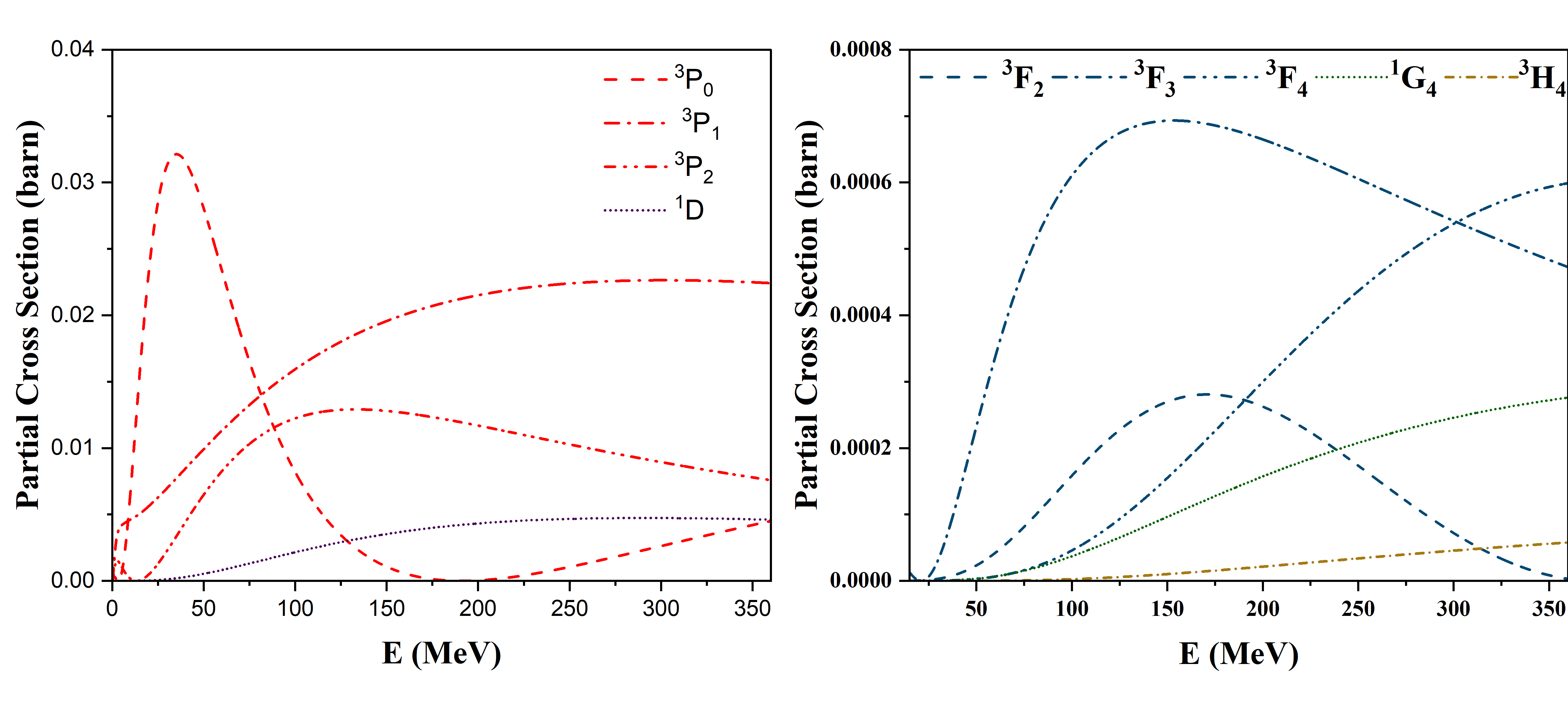}
    \caption{Partial Cross section for P, D, F, G and H state as a function of energy}
    \label{fig:pdfgh}
\end{figure}

\end{document}